\documentclass[10pt]{article}

\usepackage{amsmath}
\usepackage{amssymb}

\usepackage{graphicx}

\usepackage{cite}

\usepackage{color} 

\usepackage[labelfont=bf,labelsep=period,justification=raggedright]{caption}

\makeatletter
\renewcommand{\@biblabel}[1]{\quad#1.}
\makeatother

\date{}

\begin{document}

\begin{flushleft}
{\Large
\textbf{Nonequilibrium Brownian motion beyond the effective temperature}
}
\\
Andrea Gnoli$^{1,2}$, 
Andrea Puglisi$^{1,2,\ast}$, 
Alessandro Sarracino$^{1,2,3}$
Angelo Vulpiani$^{1,2}$
\\
\bf{1} Istituto dei Sistemi Complessi - Consiglio Nazionale delle Ricerche, Rome, Italy
\\
\bf{2} Dipartimento di Fisica, Universit\`a ''Sapienza'', Rome, Italy
\\
\bf{3} Laboratoire de Physique Th\'eorique de la Mati\`ere Condens\'ee - Centre National de la Recherche Scientifique Unit\'e mixte de recherche 7600, Universit\'e Paris 6, Paris, France
\\
$\ast$ E-mail: andrea.puglisi@roma1.infn.it
\end{flushleft}

\section*{Abstract}
The condition of thermal equilibrium simplifies the theoretical
treatment of fluctuations as found in the celebrated Einstein's
relation between mobility and diffusivity for Brownian motion. Several
recent theories relax the hypothesis of thermal equilibrium resulting
in at least two main scenarios.  With well separated timescales, as in
aging glassy systems, equilibrium Fluctuation-Dissipation Theorem
applies at each scale with its own ``effective'' temperature. With
mixed timescales, as for example in active or granular fluids or in
turbulence, temperature is no more well-defined, the dynamical nature
of fluctuations fully emerges and a Generalized
Fluctuation-Dissipation Theorem (GFDT) applies. Here, we study
experimentally the mixed timescale regime by studying fluctuations and
linear response in the Brownian motion of a rotating intruder immersed
in a vibro-fluidized granular medium. Increasing the packing fraction,
the system is moved from a dilute single-timescale regime toward a
denser multiple-timescale stage.  Einstein's relation holds in the
former and is violated in the latter. The violation cannot be
explained in terms of effective temperatures, while the GFDT is able
to impute it to the emergence of a strong coupling between the
intruder and the surrounding fluid. Direct experimental measurements
confirm the development of spatial correlations in the system when the
density is increased.


\section*{Introduction}

Several fundamental results of statistical mechanics are obtained
under the crucial assumption of thermal equilibrium. A celebrated
example of the power of the equilibrium hypothesis is given by the
theoretical treatment of Brownian motion developed by Einstein at the
beginning of the $20$th century~\cite{einstein}. Such a hypothesis
imposes symmetry under time-reversal and offers crucial shortcuts in
computations.  For instance, in calculating the self-diffusion coefficient, thermal
equilibrium provides a simple expression for the osmotic pressure of
suspended particles.  Later, in Langevin's approach, the
simplification comes from the energy equipartition which determines
straightforwardly the mean kinetic energy of Brownian particles.  The
subsequent evolution of the linear response theory has been entirely
based upon equilibrium which is the root of the celebrated
Fluctuation-Dissipation Theorem (FDT)~\cite{FDT}. This theorem states that
whenever an equilibrium system with Hamiltonian $H$, at temperature $T$,
is perturbed in such a way that its Hamiltonian changes into $H+\Delta
H(t)$, with $\Delta H(t)=A(t) \times B(t)$ ($B$ being a state function
of the system, coupled with the external force $A(t)$), then the
mean linear response for the average time evolution of an observable $O(t)$
reads
\begin{equation} \label{fdt}
\frac{\overline{\delta O(t)}}{\delta A(s)}=\frac{1}{k_B T}\left\langle O(t)
\frac{d}{ds} B(s)\right\rangle,
\end{equation}
where $\langle O\rangle$ is the average of the unperturbed system, 
$\overline{O}$ is the perturbed one and $k_B$ is the Boltzmann's constant. If the system is invariant for
time-translations, in Eq.~\eqref{fdt} the times $t$ and $s$ may be
replaced by $t-s$ and $0$, respectively.
The FDT is a powerful tool which allows the computation
of the effect of small external forces, for instance all transport
coefficients, while ignoring such forces. If, for instance, the system is perturbed by an impulse $F(t)$ at time $0$, with
$F(t) = m\delta v(0) \delta(t)$ (which in the Hamiltonian appears
coupled with $-x(t)$) applied to a particle of mass $m$, position
$x(t)$ and velocity $v(t)$, the FDT reads
\begin{equation} \label{e1}
\frac{\overline{\delta v(t)}}{m\delta v(0)} = \frac{1}{k_B T}\langle v(t)v(0)
\rangle.
\end{equation}
Eq.~\eqref{e1}, time-integrated in $[0,\infty)$, returns Einstein's relation
between diffusivity and mobility.

Recently, it has been shown that even in out-of-equilibrium systems a
relation between response and spontaneous fluctuations still
exists~\cite{BPRV08,seifert} which takes a more complicated form than
the one at equilibrium.  An important instance is represented by spin
and structural glasses which, cooled below the glass transition
temperature, display an extremely slow relaxation called aging~\cite{aging}.  A
fundamental observation is that, in some cases, timescales of relevant
degrees of freedom are separated into almost perfectly isolated
classes, i.e. very fast and very slow evolutions, and an appropriate
description of the system can be formulated by introducing the concept
of ``effective'' temperature~\cite{C11}.  For instance, in several
models, it has been shown that the FDT, Eq.~\eqref{fdt}, can be
considered as a good approximation, by replacing $T$ with an effective
time-dependent temperature $T_{eff}(t,s)$ which, for large times,
assumes a thermodynamic meaning~\cite{cris}.  Experimental verifications of this
scenario have been reported~\cite{WSM06,JPPC09}.  For driven systems,
like fluids under shear, the effective temperature scenario is
expected to hold for slow energy flows, namely for slight stirring
(which corresponds to the large time limit of glassy models). In
particular, this is the case of weakly shaken ``glassy'' granular
media, with density close to jamming~\cite{MK02,ono02}.

Often in nonequilibrium systems the different timescales are not
clearly separated and the picture in terms of effective temperature
does not hold. Instances of this entanglement of scales appear in
climate and turbulence~\cite{BPRV08}, as well as among the so-called
active fluids. They include compounds of actine filaments, swarms of
bacteria, bird flocks or fish schools, assemblies of
micro-electro-mechanical systems, collective human dynamics
(pedestrians, traffic and so on)~\cite{VZ12,marchetti}.  The validity
of the concept of effective temperature in active matter is under
intense debate, with positive~\cite{LMC08,BK13} and
negative~\cite{BSL12,FM12} answers. Several general approaches to the FDT in nonequilibrium systems have been proposed
recently~\cite{LCZ05,BPRV08,BMW09}. Some of these stress the
relevance of the unperturbed statistical distribution in phase space
which, as a rule, includes both non-Gibbsian contributions and
dynamical couplings with usually no role in the FDT at
equilibrium. Others, connecting the FDT to entropy production and to
the so-called \emph{dynamical activity}, give more importance to the
statistical distribution in path space and its simmetries under
time-reversal~\cite{BMW09}.  The formulation of the FDT used in this
paper~\cite{BPRV08}, called Generalized FDT (GFDT), for the sake of
simplicity here expressed in terms of an impulsive force and velocity
measurement, reads
\begin{equation} \label{gfdt}
\frac{\overline{\delta v(t)}}{\delta v(0)} = -\left\langle v(t)
\left.\frac{\partial \ln P(v,...)}{\partial v}\right|_{t=0}\right\rangle,
\end{equation}
where $P(v,...)$ is the unperturbed steady state distribution in the whole phase space,
involving all the relevant variables, that is not
only the perturbed particle but all the surrounding particles of the
fluid. It is clear that at equilibrium, where $P(v,...) \propto
e^{-H(v,...)/(k_B T)}$, the impulsive form of Einstein's relation,
Eq.~\eqref{e1}, is recovered.

\section*{Results and Discussion}

A paradigmatic case in which Eq.~\eqref{gfdt} can be tested is that of
strongly fluidized granular media~\cite{JNB96} for which the overall
effect of the energy injection mechanism and the presence of energy
exchanges on different space- or time-scales can induce complex
behaviors. In such systems, interactions among particles are
dissipative due to the energy loss during the collisions and an
external source is necessary in order to sustain a fluid stationary
state. A strict analogy with simple Brownian motion was shown in a
previous work analysing the rotational motion of a torsion oscillator
immersed in a dense granular fluid~\cite{AMBLN03}. By measuring noise
and susceptibility in the system, the authors found that an effective
description can be obtained within the equilibrium formalism and
showed that the shaken granular medium acts as a ``thermal'' bath
satisfying the FDT. Here, we consider a new experiment, described in
Figure~1 (see section Methods for further details), where a rotating
wheel performs granular Brownian motion immersed in a shaken granular
media~\cite{GPDPGSP13} and is weakly perturbed by the impulsive action
of a small motor. The motor is switched on for a very short lapse of
time, and exerts -- at an arbitrary time set to $0$ -- a variation of the wheel's angular velocity $\delta
\omega(0)$. We explore the range of low and medium densities (up to a
maximum of $15\%$ of packing fraction) in order to assess multiscale
regimes not considered previously~\cite{AMBLN03}.

\subsection*{Linear response}

The measurements of interest in our experiment are the response of the
angular velocity $\omega$ of the wheel to the perturbation,
$R(t)\equiv \overline{\delta \omega(t)}/\delta \omega(0)$, and
the time-correlations of the unperturbed signal $\omega(t)$, \emph{in
  primis} the classical auto-correlation $C(t)=\langle
\omega(t)\omega(0)\rangle$.  In Fig. 2, for different values of the
gas density, we show the results for $R(t)$ superimposed to
$C(t)/C(0)$. In the dilute limit, panel (a), correlations and
response functions are very close, so that $R(t) \approx C(t)/C(0)$
with slight departures which we ascribe to the large noise of the
response signal. This observation, even more compelling in the inset
of Fig. 2a showing a parametric plot $R$ \emph{vs} $C$, is equivalent
to verifying Einstein's relation, Eq.~\eqref{e1}. Note that,
normalizing the response function, the measurement of a
proportionality factor $1/C(0)=1/\langle \omega^2 \rangle$ is
inevitable even if equipartition is not satisfied (indeed, $I \langle
\omega^2 \rangle < T_g$ because of inelastic collisions, where $I$ is
the momentum of inertia of the wheel and $T_g$ is the granular kinetic
temperature). The fact that a Brownian particle suspended in a dilute
granular fluid behaves as if it were at equilibrium has been observed
before~\cite{SVCP10}: the separation of scales guaranteed by
diluteness allows the granular gas to be considered almost independent
upon the dynamics of the wheel; such a decoupling implies that each
inelastic collision of the wheel with a gas particle may be understood
as an elastic collision with different effective masses.

For higher values of the gas density, panels (b) and (c) of Fig. 2, the
scenario changes considerably.  Here, the dynamics of the tracer and
of the gas have to be considered coupled, leading to significant
deviations between response and correlation functions. Einstein's
relation is no more satisfied at packing fractions greater or equal to
$10\%$. The comparison between panels (b) and (c) of Figure~2,
better visible in their insets, shows that the amount of violation
increases with the packing fraction.

The GFDT discussed above, Eq.~\eqref{gfdt}, accounts for all the
observations of Fig. 2. In our case it reads
\begin{equation} 
R(t)=-\left\langle \omega(t)\left. \frac{\partial \ln
P(\omega,\{v_i\})}{\partial \omega}\right|_{t=0}\right\rangle.
\label{gfdt2}
\end{equation}
The static properties of the system are fully described by the joint
probability density function (PDF) $P(\omega,\{v_i\})$ of $\omega$ and
of the gas particle velocities $v_i$, with $i=1,\ldots,N$.  In
Figure~3, we show the PDF $P_\omega(\omega)$ of the angular velocity of
the rotator for different gas densities. It corresponds to the {\em
  marginalized} $P_\omega(\omega)=\int dv_1\ldots dv_N
P(\omega,\{v_i\})$ of the joint PDF. The determination of the complete
joint PDF is out of the scope of our experimental apparatus. However,
steps in this direction are discussed at the end of the
paper. Deviations from a Gaussian, in the PDF of the rotator's
angular velocity, appear at all densities.  Such discrepancies include
a slightly enhanced peak at small velocity, due to the presence of dry
friction~\cite{H05}, as well as tails slightly larger than Gaussian at
high velocities, whose origin is likely to be the inelasticity of
collisions. A good fit of $P_\omega(\omega)$ may be obtained in the
form of
\begin{equation} \label{fit}
-\ln P_\omega(\omega) = a \omega^2 + b |\omega| + c \omega^4 + const.
\end{equation}
The parameters of the fits in the three cases are $a=0.00727,
b=0.00976, c=-9\cdot 10^{-7}$ for $N=280$; $a=0.0165, b=0.0249,
c=-6\cdot 10^{-6}$ for $N=560$; $a=0.024, b=0.058, c=-1.5\cdot
10^{-5}$ for $N=840$. Units for $a$, $b$ and $c$ are $1/s^2$, $1/s$
and $1/s^4$ respectively. Negative values for coefficient $c$ are of
course non-physical at very high velocities, however they give reason
of a good fit in the observable range; one may imagine that further
corrections at higher order (irrelevant in this study) are present.

Assuming a factorization among $\omega$ and $\{v_i\}$,
i.e. $P(\omega,\{v_i\})=P_\omega(\omega)P_v(\{v_i\})$, one has
\begin{equation} \label{decoupl}
\frac{\partial \ln P_\omega(\omega)}{\partial  \omega} = \frac{\partial \ln
P(\omega,\{v_i\})}{\partial \omega},
\end{equation}
that used with~\eqref{fit} for the GFDT gives $R_G(t) =
2aC(t)+bC_1(t)+4 cC_2(t)$, with $C_1(t)=\langle
\omega(t)\textrm{sign}[\omega(0)]\rangle$ and $C_2(t)=\langle
\omega(t)\omega^3(0)\rangle$.  The non-Gaussian form of the PDF
clearly modifies the relation between response and correlation.
However, as already observed in molecular dynamics
simulations~\cite{PBV07}, it may happen that the ``extra'' correlators
$C_1(t)$ and $C_2(t)$ coming from non-Gaussianity do not deviate
substantially (once normalized to be $1$ at the origin, $t=0$) from
the velocity-velocity correlation function $C(t)/C(0)$. Our experiment shows clearly, see Fig.~2 (in particular the curves with green diamonds), 
that in all cases (dilute and more dense) the correction
induced only by non-Gaussian terms is very small, i.e. $R_G(t) \approx
C(t)/C(0)$. The first implication of this is that our experiment is in
agreement with the GFDT in the dilute case (Fig.~2a). The
second implication is that the breakdown of Einstein's relation can
only be imputed to the failure of assumption~\eqref{decoupl} in the
more dense cases (Fig.~2b and~2c).

\subsection*{Coupling with the fluid}

The emergence of the relevance of coupling between wheel and fluid,
going from the dilute case to the dense one, already appears in the
study of autocorrelations functions. At low packing fraction, the
shape of $C(t)$ is dominated by a single exponential decay with an
almost negligible negative part which displays a power law decay at
large times. The presence of a time interval with $C(t)<0$ and the
final power law decay become more and more important as the density is
increased.  In Figure~\ref{loglogtail}, we plot $|C(t)|/C(0)$ in
log-log scale for different densities. At each density a time $t^*$
exists where $C(t)$ change sign, from $C(t)>0$ to $C(t)<0$, well
evident in Fig.~\ref{loglogtail} as a sudden change of the
derivative. The negative region is reminiscent of backscattering
phenomenon and characterizes also equilibrium molecular fluids with
memory effects arising at high density.  The slow final decay $\sim
t^{-\alpha}$ with $\alpha \approx 1$ is analogous to the phenomenon of
long-time tails whose existence is acknowledged in granular
systems~\cite{OK07} and is due to the coupling of the tracer's density
with the fluid's shear flow~\cite{FAZ09}. Both the negative region and
the power-law decay become more and more relevant as the density is
increased.

Both these features imply the existence of more than one
time-scale. In a molecular fluid at equilibrium, however, even when
$C(t)$ shows such a non-trivial behavior, particles velocities remain
statistically independent as a consequence of $P \sim e^{-H/(k_B T)}$,
so that Eq.~\eqref{decoupl} holds and Einstein's relation remains
satisfied.  Out of equilibrium, on the other side, the coupling
between rotator and particles, suggested by the multiscale behavior,
induce velocity correlations among different degrees of
freedom~\cite{SVGP10}. Such an entangled joint PDF can no more be
replaced by the marginalized $P_\omega(\omega)$, in Eq.~(\ref{gfdt2}):
its ultimate consequence is the breakdown of Einstein's relation. In
our experiment the presence of slowly-decaying correlations is present
at all values of the density. However, 
 we point out that such correlations intensify with the increase of
density.  As a consequence, it is plausible that the observed violation of
Einstein's relation is due to the appearance of internal correlations
that becomes important when the density is increased. This hypotesis
is also supported by the study, discussed in the following, of
rotator-gas correlations.

In order to find an explicit form for the correlation functions
appearing in the GFDT, Eq.~\eqref{gfdt2}, it is necessary to
understand the role of the relevant degrees of freedom coupled with
$\omega$.  In certain cases, it has been shown that the dominant
contribution of this coupling consists in a ``hydrodynamic'' velocity
field (related to gas particles surrounding the
wheel)~\cite{PBV07,VBPV09,SVGP10}. The correlation between the rotator
and such a local velocity field implies a correction to
Eq.~\eqref{decoupl} and, therefore, to Einstein's relation; the physical
meaning is the emergence in the dynamics of the rotator of another
timescale related to the typical relaxation time of the local field fluctuations. In
Fig.~5, we have verified the existence in the dense regime of such a
coupling by plotting the cross-correlation $\langle \Omega(t)
\omega(0) \rangle$, where $\Omega(t)= \frac{1}{N}\sum_{i=1}^{N}
\Omega_i(t)$ and $\Omega_i(t)={\bf r}_i(t) \times {\bf
  v}_i(t)/[r(t)^2]$ is the angular velocity of particle $i$ at
position ${\bf r}_i$ relative to the center of rotation. The same measurement (properly rescaled)
reveals a much less evident coupling in the dilute configuration. This
is a strong evidence that correlations between $\omega$ and $v_i$ are
relevant and, therefore, Eq.~\eqref{decoupl} does not hold. The fair
coincidence in time of the maximum of the cross-correlation with the
region of maximum violation of Einstein's relation ($\sim 0.05$
seconds) corroborates our argument. The attempt to fit responses and
correlations through a simple model~\cite{SVGP10} with two linearly coupled
stochastic variables ($\omega$ and $\Omega$) had negative results: the behavior of our
experiment is rather complex and it is difficult even providing a
conjecture for the functional shape of $P(\omega,\Omega)$.  Indeed,
the slow decay of autocorrelations at large times is a phenomenon
which is incompatible with a simple linear model. There is the need of
a more refined kinetic theory, possibly in terms of perturbative
expansions, such as the Mode Coupling Theory~\cite{mct}, and tailored
to our two-component system (wheel and granular gas) characterised by
two different, yet coupled, kinetic temperatures.

\section*{Methods}

The granular medium, made of $N$ non-magnetic steel beads, diameter 4
mm and mass $m=0.27$ g, is housed in a polymethyl-methacrylate (PMMA)
cylinder (diameter 90 mm) with a conical-shaped floor. A fixed holder
encloses a miniaturized angular encoder (model AEDA-3300 by Avago
Technologies). The encoder, which also supports the rotator (see
below), provides high resolution measurements (up to 80,000
division/revolution at the maximum rate of 20 kHz) of the rotator
position.  The encoder is used at one half of its maximum sensitivity
that corresponds to a resolution of $0.00016$ rads, with an
acquisition rate of $200$ samples per second. The cylinder is vibrated
by an electrodynamic shaker (model V450 by LDS Test \& Measurement)
fed by a sinusoidal excitation. An accelerometer measures the actual
acceleration induced to the system. A high-speed camera (EoSens CL by
Mikrotron) tracks single beads at $200$ frames per second, in order to
measure their velocity: uncertainty in the determination of the centre
of mass of spheres is estimated to be $\sim 0.05$
mm~\cite{GPDPGSP13}. A PMMA rectangular parallelepiped, termed
``wheel'' in the paper, of height $h=15$ mm and rectangular base with
dimensions $34 \times 6$ mm$^2$ is suspended, by a rod through a small
hole in the top face, to the angular encoder that records the wheel's
angle. The momentum of inertia of the free rotator (cylinder plus rod)
is $I_{rot}=353$ g mm$^2$. The setup is similar to that used in
Ref.~\cite{GPDPGSP13} with the addition of a miniaturized dc motor
(model 108-105 from Precision Microdrives) connected, through a couple
of gears, to the rotation axis of the wheel. We have not measured the
total momentum of inertia $I>I_{rot}$ of the rotator coupled to the
motor. The motor is driven by sharp rectangular electrical pulses
provided by the acquisition board (model NI USB-6353 from National
Instruments) through a simple voltage buffer circuit. The effect of
the pulses is to perturb the rotator's velocity that is the variable
taken in consideration here. We use 2 ms long and 5 V high pulses
provided to the motor every second. We have verified that both the
response and correlation functions take less than one second to go to
zero, i.e. all perturbations can be considered independent. We have
also directly checked the linear response regime. In order to have
clean response measurements, we performed 70 hours long
experiments. The acquisition rate of the system is set at 200 Hz. We
use three different gas densities ($\sim 5$\%, $\sim 10$\% and $\sim
15$\% of the total volume) varying the number of beads ($N=280$, 560
and 840, respectively). By a careful particle tracking
procedure~\cite{GPDPGSP13} we can measure one of the horizontal
components (on the plane) of the particles' velocity $v$, which gives
access to the so-called granular temperature $T_g = m \langle v^2
\rangle$.  Our choice to employ a wheel which is free to rotate around
a fixed (vertical) axis, instead of a torsion
oscillator~\cite{AMBLN03}, is only motivated by simplicity of
realization. Of course, such different choice is irrelevant for the
regime of linear response.

At the currently used maximum acceleration ($24.3$ in units of gravity
acceleration), the typical horizontal velocity $v_0=\sqrt{T_g/m}$ of
particles goes from $\approx 145$ mm/s at the maximum density
($N=840$) to $\approx 250$ mm/s at the minimum one
($N=280$). Estimates of the particle-particle mean free path give
$\approx 40$ mm for the more dilute experiment and $\approx 14$ mm for
the more dense one. The estimate for the mean free time for
particle-particle collisions goes from $\approx 0.1$ to $\approx 0.17$
seconds, for the more dense and the more dilute experiment
respectively. The mean free time of the rotator (which does not
distinguish between different particles) is $\approx 0.011$ seconds in
the most dilute experiment and $\approx 0.007$ seconds in the most
dense one.

\section*{Acknowledgments}
We would like to thank MD. Deen for technical support and A. Petri for
useful comments on the manuscript. The authors acknowledge the support
of the Italian MIUR under the grants: FIRB-IDEAS n. RBID08Z9JE. AP and
AV acknowledge the support of the Italian MIUR under the grant PRIN
n. 2009PYYZM5.


\newpage
\section*{Figure Legends}

\begin{figure}[!ht]
\begin{center}
\includegraphics[width=9cm]{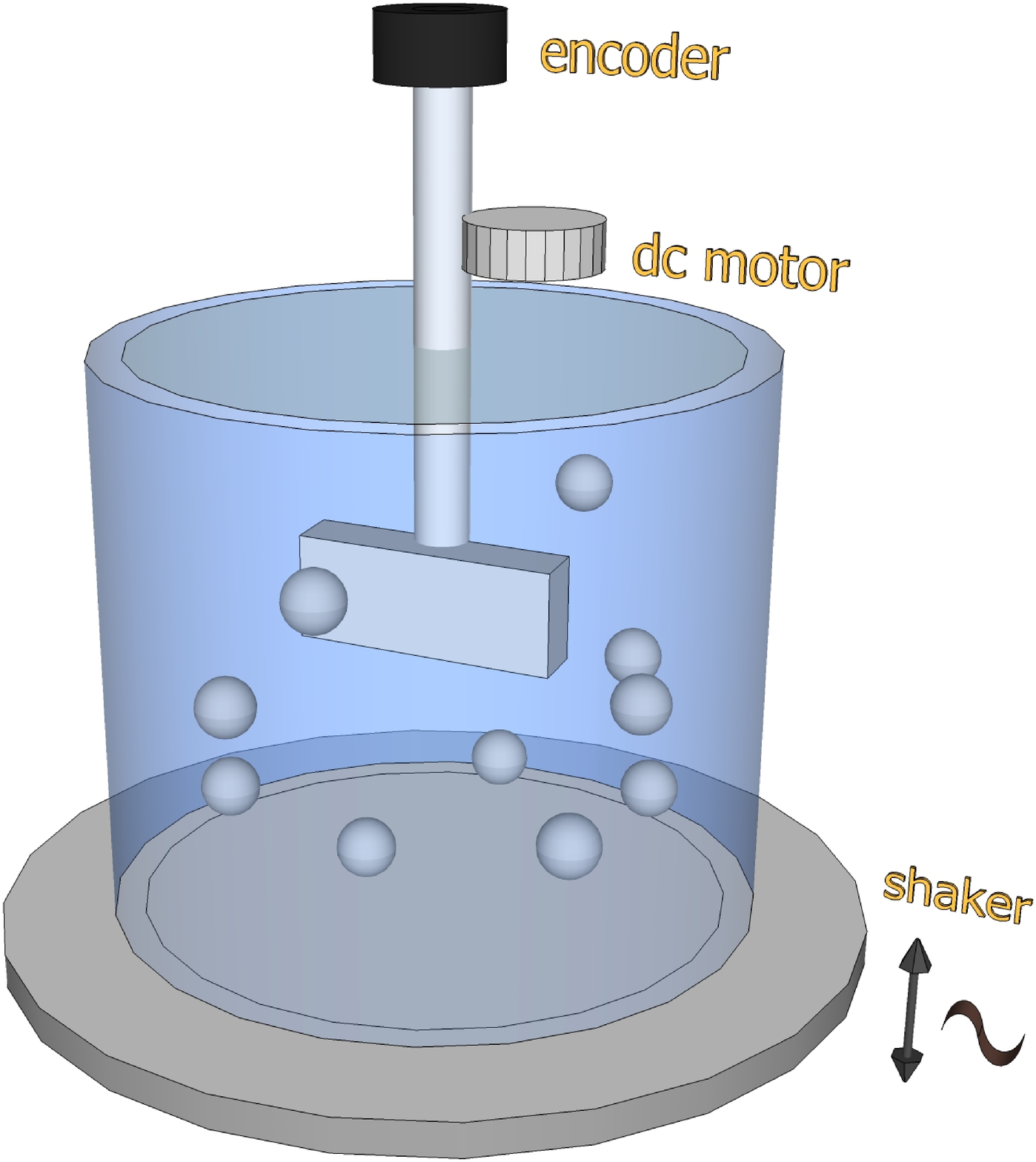}
\end{center}
\caption{{\bf Experimental setup} A sketch of the setup illustrates the essential
  components. A wheel rotating around a fixed axis is suspended in a
  cylindrical cell containing steel spheres. The cell is shaken in
  order to fluidize the material and obtain a granular gas. The wheel
  performs a Brownian-like dynamics, randomly excited by collisions
  with the spheres. A small motor is coupled to the wheel axis, in
  order to apply an external impulsive perturbation. An angular
  encoder reads the angular velocity of the wheel. Statistical
  properties of the velocities of the spheres are collected through a
  fast camera, placed above the system. A detailed description is
  presented in Methods section.}
\label{setup}
\end{figure}

\begin{figure}[!ht]
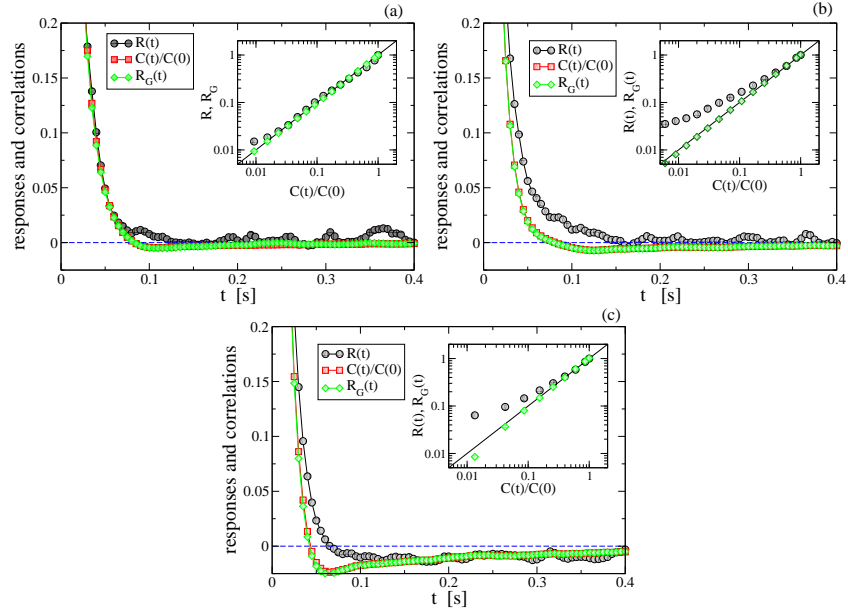

\begin{center}
\includegraphics[width=5.5cm,clip=true]{new_fig_dil.eps}
\includegraphics[width=5.5cm,clip=true]{new_fig_med.eps}
\includegraphics[width=5.5cm,clip=true]{new_fig_den.eps}
\end{center}
\caption{{\bf Response and autocorrelation}. Response function $R(t)$ (black circles), rescaled velocity
autocorrelation $C(t)/C(0)$ (red squares), and GFDT response with the
factorization assumption, Eq.~\eqref{decoupl},
$R_G(t)=2aC(t)+bC_1(t)+4cC_2(t)$ (green diamonds) for $N=280$ (a),
$N=560$ (b) and $N=840$ (c), that is packing fractions $5\%$, $10\%$
and $15\%$, respectively. In the inset the parametric plot $R, R_G$
\emph{vs} $C$, in the region where $C$ is positive and monotonously
decreasing, is plotted in log-log scale.  In the densest cases, $R(t)$
and $C(t)$ behave very differently and Einstein's relation is
significantly violated.}
\label{fig_gfdt}
\end{figure}

\begin{figure}[!ht]
\begin{center}
\includegraphics[width=8cm,clip=true]{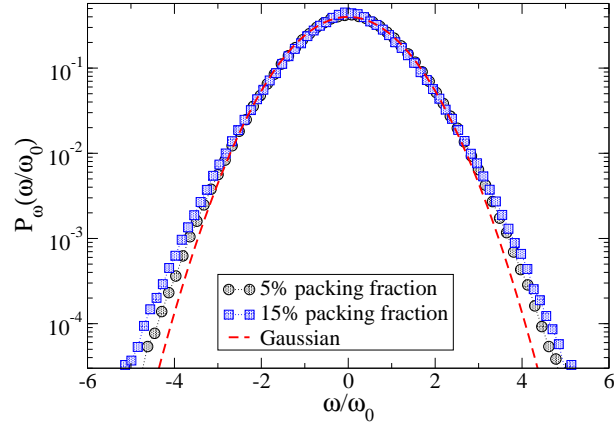} 
\end{center}
\caption{{\bf Velocity distributions}. PDF of the rotator's angular
velocity rescaled by $\omega_0=\sqrt{\langle \omega^2\rangle}$ for low
(black circles, $\omega_0=7.7$ rad/s) and high (blue squares,
$\omega_0=4.1$ rad/s) densities. The red dashed line shows a Gaussian
fit for comparison.}
\label{pdf}
\end{figure}

\begin{figure}[!ht]
\begin{center}
\includegraphics[width=9cm,clip=true]{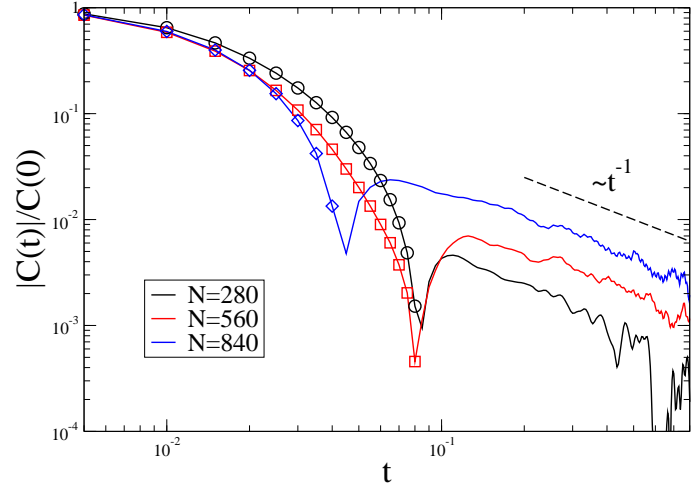}
\end{center}
\caption{{\bf Long tails}. Absolute value of autocorrelations in log-log
scale (symbols denote positive values) for different densities. }
\label{loglogtail} 
\end{figure}

\begin{figure}[!ht]
\begin{center}
\includegraphics[width=8cm,clip=true]{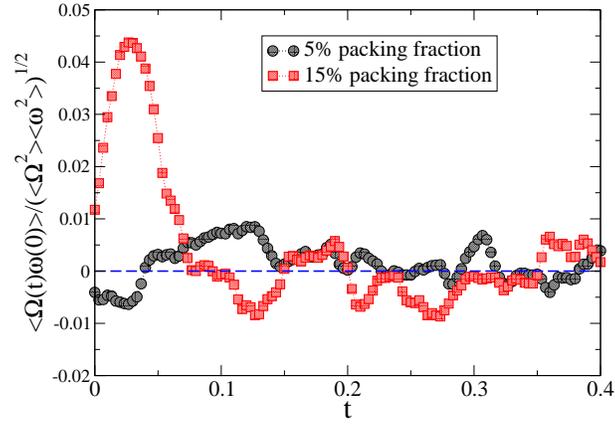}
\end{center}
\caption{{\bf Coupling with the gas}. Correlation between the angular velocity of the probe
 $\omega(t)$ and the average angular velocity
of the fluid $\Omega(t)$ (see text for definition) for the most dilute and the most
dense experiments.}
\label{coupling}
\end{figure}


\end{document}